\documentclass[aps,twocolumn,superscriptaddress,amsmath,amssymb]{revtex4}

\usepackage{amsmath,amssymb}
\usepackage{graphicx}
\usepackage{dcolumn}
\usepackage{bm}

\begin{document}

\title{In-situ electron-beam lithography of deterministic single-quantum-dot mesa-structures using low-temperature cathodoluminescence spectroscopy}

\author{M.~Gschrey}
\author{F. Gericke}
\author{A.~Sch\"u{\ss}ler}
\author{R.~Schmidt}
\author{J.-H.~Schulze}
\author{T.~Heindel}
\author{S.~Rodt}
\author{A.~Strittmatter}
\author{S.~Reitzenstein}
\email{stephan.reitzenstein@physik.tu-berlin.de}
\affiliation{Institut f\"{u}r Festk\"{o}rperphysik, Technische Universit{\"a}t Berlin, Hardenbergstra{\ss}e 36, D-10623 Berlin, Germany}

\date{\today}

\begin{abstract}

We report on the deterministic fabrication of  sub-$\mu$m mesa-structures containing single quantum dots (QDs) by in-situ electron-beam lithography. The fabrication method is based on a two-step lithography process: After detecting the position and spectral features of single InGaAs QDs by cathodo-liminescence (CL) spectroscopy, circular sub-um mesa-structures are defined by high-resolution electron-beam lithography and subsequent etching. Micro-photoluminscence spectroscopy demonstrates the high optical quality of the single-QD mesa-structures with emission linewidths below 15$~\mu eV$ and g$^{(2)}(0) = 0.04$. Our lithography method has an alignment precision better than 100~nm which paves the way for a fully-deterministic device technology using in-situ CL lithography.

\end{abstract}


\maketitle

The development of optical quantum devices based on single-photon emitters such as semiconductor quantum dots (QDs) embedded into microcavities has 
revolutionized the field of quantum optics in solid-state physics~\cite{Shi07,Rei12}. Initially, and to a large extent even today, such devices rely on the selection of 
statistically grown QDs in a random process. As a result, the yield of functional devices is very low, typically below 1\%, which hinders the further development
towards a practical technology platform. Moreover, the statistical spatial and spectral coupling of self-assembled QDs to the optical modes in microcavity systems impedes systematic studies of light-matter interaction effects. For these reasons, huge efforts have been directed towards a deterministic device technology. One route to achieve 
this goal is based on the site-controlled growth of QDs and their integration into nanophotonic devices, and enormous progress has been achieved in this field in recent years~\cite{Sch07,Sch08,Sch09,Ski11,Jon12,Str12,Unr12}. However, site-controlled QDs still suffer from a degraded optical quality in terms of emission linewidths and quantum efficiency as compared to standard QDs based on the Stranski-Krastanow growth mode~\cite{Alb10}. In addition, their emission 
energy, and thus the spectral matching in QD-microcavities, is hardly controlled during growth. In an alternative and fully deterministic approach, in-situ optical lithography 
has been established to select the position and emission energy of target QDs before defining spatially and spectrally matched microcavity structures~\cite{Dou08}. 
This very attractive technology platform has the drawback that optical lithography has limited lateral resolution of a few hundred nm at best.

\begin{figure}[h]
\centering
\includegraphics[width=0.45\textwidth]{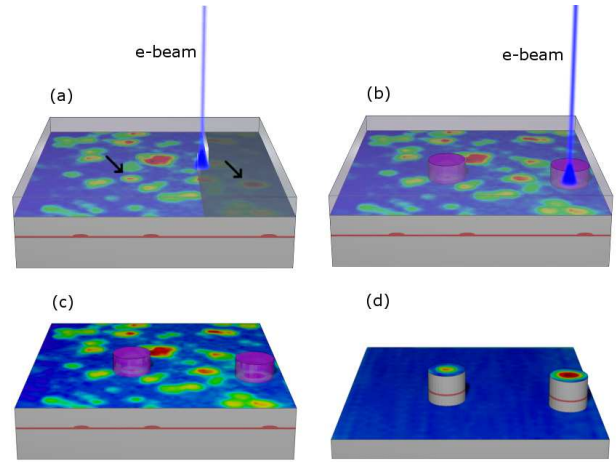}
\caption{(Color online) Schematic view of the CLL process flow. After selecting target QDs by CL-mapping (a), sub-$\mu$m mesa structures are written
into the PMMA resist (b) which acts as a negative resist by inverting at the large doses used in the lithography step. After developing the resist (c) the electron-beam defined 
structures are transferred into the semiconductor material by dry chemical etching which results in single-QD mesa structures.}
\label{fig:figure1}
\end{figure}

In contrast, electron-beam lithography (EBL) can achieve a resolution of a few nm which is a prerequisite for the realization of advanced nanodevices. A combination of EBL and CL is very promising for the site-selective exposure of the resist with respect to the position of optically active nanostructures like, e.g., QDs. In the following this technique will be dubbed cathodoluminescence lithography (CLL). Such an approach has been used to deterministically contact single nanowires in \cite{Don10}, but without the need for a spectral selection of the wires. Extending CLL with respect to nanophotonic devices could strongly boost the development of QD micro- and nano-cavities by enabling the fabrication of fully-deterministic structures with nm accuracy and a very-flexible device design in combination with spectral matching.

In this letter we report on a deterministic nanophotonic device technology based on low-temperature CLL. Our technology platform combines the advantages 
of CL spectroscopy and high-resolution electron-beam writing into a unique lithography method for nanophotonic devices. To demonstrate the power of this
method we take advantage of the high spatial resolution of CL to precisely determine the location of target QDs and to fabricate nm-scaled mesa structures 
spatially aligned to the selected QDs. Optical studies on the patterned samples prove the high yield and accuracy of this lithography technique and show
its suitability for the fabrication of high-quality nanophotonic structures. Our CLL method has high potential to pave the way for a new generation 
of quantum devices where nanoscale cavity structures can be aligned to quantum emitters with very high precision in order to realize, e.g., non-classical
light sources or a single QD-laser for on-chip optical networks. Moreover, it could also be used for the deterministic fabrication of nanoelectronic devices
such as QD-based flash-memories relying on a precise positioning of single QDs in the active layer.

As an in-situ lithography method, CLL is performed without the need of alignment markers or a transfer of the sample into another system
for electron-beam writing. This leads to a potentially more reliable and simpler process flow as compared to other electron-beam lithography approaches using 
for instance optical spectroscopy to identify the position of QDs relative to alignment markers, which allows one to retrieve their positions in a subsequent 
electron-beam lithography step after transferring the sample~\cite{Toj13}. The process flow of our approach is depicted in figure~\ref{fig:figure1}. After spin coating the 
QD-sample with standard electron-beam resist (polymethylmethacrylat, 950K PMMA) we perform CL spectroscopy to build a luminescence map from individual QDs to measure their 
emission spectra and to determine their spatial positions with an accuracy of less than 50~nm (Fig.~\ref{fig:figure1}(a)). In this step it is crucial to 
ensure an homogenous exposure without inverting the resist. In accordance with \cite{Don10}, the resist was found to sustain low temperatures very well and to be less sensitive for electron doses as compared to room temperature. Once suitable QDs have been identified in the resulting CL map (indicated by black arrows), circular patterns that are spatially aligned to the selected QDs are written into the resist (Fig.~\ref{fig:figure1}(b)), where the mesa-diameter is controlled precisely by the applied dose. After developing the resist (Fig.~\ref{fig:figure1}(c)) the remaining PMMA acts as etching mask in a subsequent plasma etching step (Fig.~\ref{fig:figure1}(d)) which results in single-QD nanostructures as will be shown below.

\begin{figure}[h]
\centering
\includegraphics[width=0.45\textwidth]{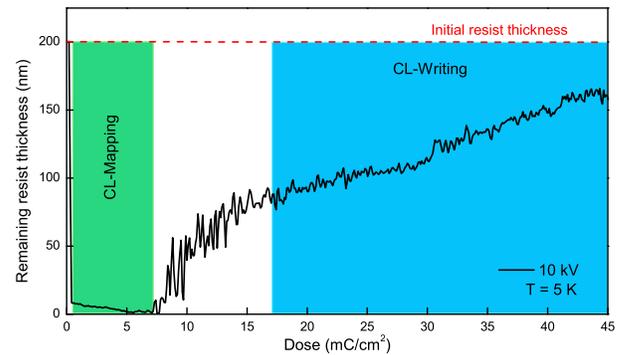}
\caption{(Color online) Characteristics of the e-beam resist (PMMA) in the CLL-process at low temperature (5~K) and an acceleration voltage of 10~kV. 
The remaining thickness of the resist after development is depicted as a function of the exposure dose. The resist with an initial thickness of 200~nm acts as 
positive resist up to about 8~mC/cm$^2$ and has negative characteristics for larger doses by inverting. This feature is exploited in CLL where CL-mapping is 
performed at intermediate doses as indicated in green and electron-beam writing is done at larger doses of up to 45~mC/cm$^2$ as indicated in blue.}
\label{fig:figure2}
\end{figure}

Most critical in CLL is the careful choice of the CL-scanning parameters in the QD-selection process and of the exposure parameters in the subsequent lithography step. 
In fact, during CL-scanning it is necessary to achieve a good balance between a low exposure dose (as a function of beam current, acceleration voltage and integration time per spectrum) and 
sufficiently high CL signal. It turned out that PMMA, which is a positive resist, cannot be used in the normal mode because the lowest possible dose ($<$ 500 $\mu$C/cm$^2$) 
during CL scanning already fully cracks the polymer and would lead to a complete removal of the exposed resist in the development step. This is illustrated in Fig.~\ref{fig:figure2} 
where the remaining thickness after cold developing (IPA, -35$\,^{\circ}\mathrm{C}$) an exposed layer of 200 nm PMMA on a GaAs sample is shown after scanning with an acceleration voltage of 10 keV and increasing exposure times. The range of resulting exposure doses (0.5 - 7  mC/cm$^2$) in which the CL-mapping is performed is indicated in green. For larger doses the resist becomes inverted and acts as a negative resist, so that the remaining thickness increases up to about 150~nm for a dose of 40~mC/cm$^2$. This is exploited in the
lithography step which is performed for doses in the range of 14 - 45~mC/cm$^2$ as indicated in blue.

\begin{figure}[h]
\centering
\includegraphics[width=0.45\textwidth]{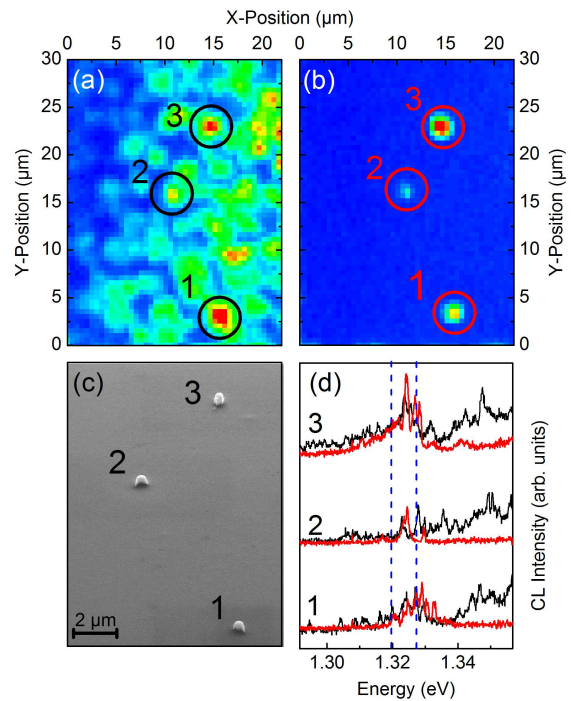}
\caption{(Color online) Low-temperature CL-intensity maps (linear scale) obtained (a) during CL spectroscopy and (b) after the processing of single-QD mesa structures for an acceleration voltage of 10 kV and a beam-current of 0.5 nA. The QDs selected during the CL-mapping (labelled 1-3) are integrated deterministically into structures with diameters below 500~nm. (c) High-resolution SEM image recorded under an angle of 60$^\circ$ of the processed sample. (d) CL-emission spectra of the selected QDs recorded during CL-mapping (black) and on the final mesa structures (red). The blue, dashed lines indicate the spectral window, which was used for the 2D-fitting routine to determine the QD position and for which the integrated intensity is displayed in (a) and (b).}
\label{fig:figure3}
\end{figure}

Our CLL method is demonstrated by fabricating circular sub-$\mu$m mesa structures aligned to pre-selected InGaAs QDs. The sample was grown by metal-organic chemical vapor depositon in an AIXTRON 200/4 machine on a 3'' GaAs (001) wafer using standard precursors. First, 500 nm of GaAs were deposited followed by 30~nm of AlGaAs as diffusion barrier for charge carriers and 150~nm of GaAs. The self-organized QDs were formed at a temperature of 500$\,^{\circ}\mathrm{C} $ during a growth interuption of 35~s. Finally the QDs were capped by 150~nm of GaAs, 27~nm of AlGaAs as second diffusion barrier and 5 nm GaAs as top layer. Due to intrinsic carbon doping during growth a p-type background doping of $\approx 1 \cdot 10^{17}$~cm$^{-3}$ is present in the sample. For AFM measurements a second layer of QDs was added on top of the sample with the same growth conditions as the first layer. At helium temperatures the QDs emit at a wavelength of around 950 nm. The QD density was estimated to be about 1$\cdot$10$^9$ cm$^{-2}$. After the sample growth we used spin coating to deposit an homogenous, 200 nm thick layer of PMMA on the sample. Then a 5 mm $\times$ 5 mm sample was transferred into the CL-system in order to perform CL-mapping for the selection of suitable QDs. A typical CL-map of a 22 $\mu$m $\times$ 30 $\mu$m sample area is shown in Fig.~\ref{fig:figure3} (a), clearly marking the positions of single QDs by bright spots. In this example, we selected three free-standing QDs labeled 1 to 3 for in-situ lithography. The corresponding electron-beam writing was performed directly after the CL-mapping without moving the sample. The position of each of the QDs was precisely determined by applying a 2D-fitting routine to the intensity distribution in the CL-map for the spectral window depicted by the blue, dashed line in Fig.~\ref{fig:figure3} (d). At each position, a circular pattern of nominal 500~nm in diameter was written. Afterwards the resist was cold developed and the 
inverted resist above the chosen QDs was used as an etch mask in the subsequent etching step using an ICP-RIE plasma (BCl$_2$ + Cl$_2$ +Ar) etcher. Figure~\ref{fig:figure3} (c) shows a high resolution SEM image of the fully processed mesas. In order to demonstrate the deterministic integration of single QDs into sub-$\mu$m mesa structures, we performed 
another CL-mapping on the processed sample. The corresponding CL-intensity map is depicted in Fig.~\ref{fig:figure3} (b) and demonstrates that all three mesa structures with pre-selected QDs are optically active. Associated CL-spectra recorded during the mapping step (black curves) and after etching (red curves) are shown in Fig.~\ref{fig:figure3} (d).  The single QD linewidths of about 0.9 meV are limited by the spectral resolution of the CL setup. The fact that the spectral features are reproduced by the processed sample clearly demonstrates the deterministic integration of selected QDs into the sub-$\mu$m mesa structures.

The reliability of our CLL method was checked by fabricating several circular mesa structures with diameters of 800 nm, 450 nm and 350 nm. For the largest diameter 89\%, i.e. 38 out of 43 mesas, of the pre-selected QDs showed unaltered luminescence as compared to measurements before fabrication of the mesas. The 450 nm-sized structures had such a yield of 67\% and a yield of 50\% was achieved for the 350 nm-sized mesas. Here we have to take into account that QDs implemented in small mesa structures are influenced by fluctuating charge distributions at defects in the dot's environment or at the lateral sidewalls, which results in spectral diffusion \cite{Turck2002} and increased emission linewidths \cite{Bay02}. Hence, further optimization of the etch process might increase our yield. 
The positioning accuracy was checked by reference samples as follows: A metallic shadow mask, fabricated with the aid of dispersed 100~nm polystyrol nano-spheres, was used as a spotty luminescence grid placed on top of a quantum-well sample. By CLL cross-markers were fabricated around the luminescence spots. The accuracy of aligning the cross-markers relative to the mask openings was better than 100 nm as confirmed by SEM measurements.

\begin{figure}[h]
\centering
\includegraphics[width=0.45\textwidth]{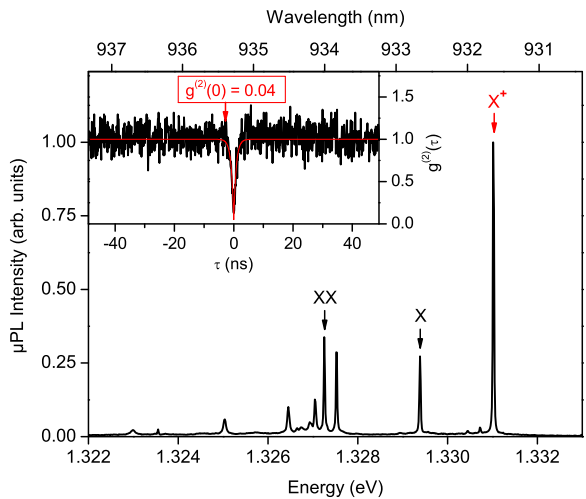}
\caption{(Color online) $\mu$PL emission spectrum of QD \#3 (T=10 K). Exciton (X), biexciton (XX) and singly charged exciton (X$^+$) emission lines are identified and the linewidths down to 9 $\mu$eV clearly reflect the high optical quality of the single-QD mesa structures. Inset: Photon auto-correlation measurement carried out on the X$^+$-emission revealing a $g^{(2)}(0)$-value as low as $0.04$.}
\label{fig:figure4}
\end{figure}

To further study the optical quality of the single-QD mesas we performed high-resolution micro-photoluminescence ($\mu$PL) spectroscopy at low temperature (T=10 K), during which the sample was optically excited by a HeNe laser. PL was collected by a 20x microscope objective with a numerical aperture of 0.4, dispersed by a monochromator and detected by a Si-charge-coupled-device camera. Figure~\ref{fig:figure4} displays a $\mu$PL-spectrum of the CLL-processed mesa containing QD \#3 (cf. Fig.~\ref{fig:figure3}) for an excitation power of 1.65~$\mu$W. By analyzing the spectral diffusion of the emission lines we could verify that all lines belong to one and the same QD \cite{Rodt05prb}. We observe narrow emission lines resulting from the recombination of the exciton (X), biexciton (XX) and singly charged exciton (X$^+$) states of the single QD. Thereby the transitions were identified by power- as well as polarization-dependent measurements. The remaining emission lines are most probably due to transitions of charged exitonic complexes as a consequence of the background doping in the sample. By fitting a Voigt profile to the emission lines of X, XX and X$^+$, at which the Gausian part was set to the spectral resolution of our setup (23 $\mu$eV), we extract Lorentzian linewidths of $\gamma_{X}=15$ $\mu$eV, $\gamma_{XX}=13$ $\mu$eV and $\gamma_{X+}=9$ $\mu$eV underlining the high optical quality of the single-QD mesa structure. Furthermore, photon autocorrelation measurements were carried out on the X$^+$-emission for an excitation power of 75 nW. The emission line was spectrally selected by a monochromator and coupled into a fiber-based Hanbury-Brown and Twiss setup (HBT). It consists of a multimode fiber-beamsplitter (50~$\mu$m core-diameter, split ration 50:50) together with two $\tau$-SPAD-50 (PicoQuant) avalanche photo diodes featuring a timing resolution down to 350 ps. In order to shift the detectors cross-talk away from the zero-delay point, a 25 m multimode fiber-delay is attached symmetrically to both arms of the HBT. Time correlated single photon counting is performed by a PicoHarp 300 module with a time bin width of 4~ps. The resulting histogram of the photon auto-correlation function g$^{(2)}(\tau)$ is shown in the inset of figure~\ref{fig:figure4}, where a time bin-width of 128 ps was chosen. A fit to the raw data yields a value of $g^{(2)}(0)=0.04$, proving the quantum nature of emission with a strong suppression of two-photon emission events. 
  
In conclusion, we have demonstrated a lithography technique which allows for the deterministic selection and processing of sample-areas containing single QDs with high yield and 
high optical quality. Our lithography approach is based on combined high-resolution CL-spectroscopy and in-situ electron-beam definition of nanophotonic structures. The 
high quality and enormous potential of the approach was demonstrated by fabricating deterministic single-QD mesa structures with sub-$\mu$m dimensions. It combines the pre-selection of quantum emitters by means of their lateral and spectral positions with the ability of patterning low-dimensional structures with nm-precision and thus will pave the way for a 
new generation of fully deterministic quantum devices with unprecedented complexity and quality.

We acknowledge support from Deutsche Forschungsgemeinschaft (DFG) through SFB 787 ''Semiconductor Nanophotonics: Materials, Models, Devices''.

\end{document}